\documentclass[%
 reprint,
superscriptaddress,
 amsmath,amssymb,
 aps,
float]{revtex4-1}

\usepackage{graphicx}
\usepackage{dcolumn}
\usepackage{bm}
\usepackage[caption=false]{subfig}
\usepackage{xcolor}

\begin{document}

\title{Switchable Damping for a One-Particle Oscillator}

\author{X. Fan}
 \email{xingfan@g.harvard.edu}
\affiliation{Department of Physics, Harvard University, Cambridge, Massachusetts 02138, USA}
 \affiliation{Center for Fundamental Physics, Northwestern University, Evanston, Illinois 60208, USA}
\author{S. E. Fayer}
 \affiliation{Center for Fundamental Physics, Northwestern University, Evanston, Illinois 60208, USA}
\author{T. G. Myers}
 \affiliation{Center for Fundamental Physics, Northwestern University, Evanston, Illinois 60208, USA}
\author{B. A. D. Sukra}
 \affiliation{Center for Fundamental Physics, Northwestern University, Evanston, Illinois 60208, USA}
\author{G. Nahal}
 \affiliation{Center for Fundamental Physics, Northwestern University, Evanston, Illinois 60208, USA}
\author{G. Gabrielse}
 \email{gerald.gabrielse@northwestern.edu}
 \affiliation{Center for Fundamental Physics, Northwestern University, Evanston, Illinois 60208, USA}
\date{\today}


\begin{abstract}
The possibility to switch the damping rate for a one-electron oscillator is demonstrated, for an electron that oscillates along the magnetic field axis in a Penning trap.  Strong axial damping can be switched on to allow this oscillation to be used for quantum nondemolition detection of the cyclotron and spin quantum state of the electron.  Weak axial damping can be switched on to circumvent the backaction of the detection motion that has limited past measurements. The newly developed switch will reduce the linewidth of the cyclotron transition of one-electron by two orders of magnitude.
\end{abstract}

\maketitle

\section{Introduction}
\label{sec:introduction}

A single isolated trapped particle is an ideal system to test predictions of the Standard Model of particle physics (SM) involving magnetic moments and charge-to-mass ratios of the electron, positron, proton and antiproton \cite{DehmeltMagneticMoment,HarvardMagneticMoment2008,HarvardMagneticMoment2011,ProtonMagneticMoment,ChargeToMassBase2015,MagneticMomentPbarBase2017}. After decades of agreement, the measured electron's magnetic moment in Bohr magnetons $\mu/\mu_B$ \cite{HarvardMagneticMoment2008,HarvardMagneticMoment2011} now disagrees with the SM prediction \cite{atomsTheoryReview2019,MullerAlpha2018} by 2.4 standard deviations. This discrepancy at the $3 \times 10^{-13}$ level has stimulated many new theoretical investigations \cite{gardner2019light,ALightComplexScalarForTheElectronAndMuonAnomalousMagneticMoments,PhysRevD.98.075011,PhysRevD.98.113002,PhysRevD.99.095034,ContributionToG2Paradisi,ContributionToG2FromALPParadisi}.  

The electron measurements were done with a single electron in the magnetic field and electrostatic quadrupole potential of a Penning trap \cite{HarvardMagneticMoment2008,HarvardMagneticMoment2011}. Quantum jump spectroscopy of fully-resolved quantum levels of the cyclotron and spin motion with an electron cooled to 0.1 K was key \cite{QuantumCyclotron}.  The detection backaction uncertainty that remains must be reduced to improve the accuracy of the measurements to investigate the intriguing discrepancy between prediction and measurement. This work is a demonstration of a 200 MHz electronic switching system at cryogenic temperatures developed to accomplish this. It will make it possible to resolve the quantum energy levels of the electron's axial motion for the first time \cite{Fan2020EvadingBackActionPRL,Fan2020EvadingBackActionPRA}. The harmonic oscillation along the magnetic field direction is a detection motion used for quantum nondemolition (QND) detection \cite{QNDScience1980,QNDReview1980, QNDreview1996} of the quantum cyclotron and spin states \cite{DehmeltMagneticBottle}. The cryogenic switching is between weak axial damping to circumvent all detection backaction, and strong axial damping to make a one-particle signal large enough to observe.   

A recent quantum calculation establishes the theoretical basis for circumventing  detector backaction \cite{Fan2020EvadingBackActionPRL,Fan2020EvadingBackActionPRA}. The calculation shows that reduced axial damping makes it possible to resolve a cyclotron resonance for each individual axial quantum state. The backaction effect is circumvented insofar as the cyclotron excitations that take place during the fraction off the time that the backaction fluctuations take the axial motion into its ground state can be resolved.

Section \ref{sec:DetectorBackaction} is a brief summary of the detector backaction that must be eliminated;  it does not repeat the derivations and calculations recently reported  \cite{Fan2020EvadingBackActionPRL,Fan2020EvadingBackActionPRA}. Sec.~\ref{sec:developments} presents the detection circuity and calculations of its effect upon the electron axial motion.  Sec.~\ref{Sec:PenningTrapMeasurement}
provides the first demonstration of the circuitry being used to change axial damping and expected lineshape. An overview of the experimental apparatus within which this circuit was used are in \cite{EfficientPositronAccumulation,atomsNewMeasurement2019,Helium3NMR2019}.  Sec.~\ref{sec:Conclusion} presents a summary and conclusions.

\section{QND Detection and Backaction}
\label{sec:DetectorBackaction}

The QND detection of the quantum cyclotron and spin states (described elsewhere in detail  \cite{Fan2020EvadingBackActionPRL,Fan2020EvadingBackActionPRA}) is summarized briefly here to provide the context needed to discuss the detector. 

The energy eigenstates of the electron in the Penning trap are direct products of independent cyclotron, spin and axial eigenstates, designated by their quantum numbers $\left|n_c, m_s, n_z\right\rangle$. For an electron cooled to $0.1$ K, only three cyclotron and spin states are populated.  Two are cyclotron ground states (with cyclotron and spin quantum numbers $n_c=0$ and $m_s = \pm 1/2$), both stable for years or more.  One is a spin down cyclotron excited state (with $m_s=-1/2$ and $n_c=1$) that has a damping time due to cavity-inhibited spontaneous emission of typically 5 seconds \cite{CylindricalPenningTrap,InhibitionLetter}.  
The energy eigenvalues for these eigenstates are 
\begin{equation}
\begin{split}
E(n_c,m_s,n_z) = 
&\hbar\omega_c\left(n_c+\tfrac{1}{2}\right)+\hbar\omega_sm_s+\hbar\omega_z\left(n_z+\tfrac{1}{2}\right)\\
+&\hbar\delta_c\left(n_c+\tfrac{1}{2}\right)\left(n_z+\tfrac{1}{2}\right) + \hbar\delta_sm_s\left(n_z+\tfrac{1}{2}\right),
\label{eq:energystate}
\end{split}
\end{equation}
where $\hbar$ is the reduced Planck constant, and the last line is due to a magnetic bottle gradient \cite{DehmeltMagneticBottle} that is introduced to couple the cyclotron and spin states to the axial states. The magnetic bottle parameters, $\delta_c$ and $\delta_s$,  are the shifts of axial frequency for one-quantum cyclotron and spin transitions, respectively \cite{DehmeltMagneticBottle,Fan2020EvadingBackActionPRL}. It is a QND coupling that does not change the energy eigenstates for the system \cite{QuantumCyclotron}.   

The magnetic bottle gradient is essential for detecting the quantum spin and cyclotron state of the electron because it couples the axial frequency to the cyclotron and spin energy.  From Eq.~(\ref{eq:energystate})  we see that axial frequency shifts,
\begin{equation}
\Delta \omega_z = (n_c+\tfrac{1}{2})\delta_c + m_s \delta_s  
\end{equation}
reveal changes in the cyclotron and spin quantum numbers. Measuring the axial frequency is thus a QND detection that does not itself change the quantum state.  Compared to the spin and cyclotron  frequencies in the best measurement \cite{HarvardMagneticMoment2008}, 
\begin{eqnarray}
&\omega_s/(2 \pi) =& 150.5 \rm{~GHz}\\
&\omega_c/(2 \pi) =& 150.3 \rm{~GHz}\\
&\omega_a/(2 \pi) =& (\omega_s -\omega_c)/(2 \pi) = 173 \rm{~MHz}
\end{eqnarray}
the corresponding magnetic bottle shifts are very small, with
\begin{eqnarray}
&\delta_s/(2 \pi) =& 3.872 \rm{~Hz}\\
&\delta_c/(2 \pi) =& 3.868 \rm{~Hz}\\
&\delta_a/(2 \pi) =& (\delta_s - \delta_c)/(2 \pi) = 0.004 \rm{~Hz}
\end{eqnarray}  
The frequencies for spin and cyclotron differs only by 1 part per 1000 since $g/2\approx1.001$. If these tiny shifts are detected, the QND coupling keeps the axial detection motion (and the transistor amplifier to which it is coupled) from altering the quantum state of the spin and cyclotron motion.  

However, the QND coupling does not prevent a backaction that shifts the cyclotron and anomaly frequencies in proportion to the axial energy.  This parallel detection backaction is an unavoidable consequence of the QND coupling. The physical reason is that axial motion through the magnetic bottle gradient changes the magnetic field in which the cyclotron and spin motions evolve. The backaction shifts from Eq.~(\ref{eq:energystate}) are 
\begin{eqnarray}
\Delta \omega_c &=& (n_z+\tfrac{1}{2})\,\delta_c \label{eq:CyclotronFrequencyShift}\\
\Delta \omega_a &=&  (n_z+\tfrac{1}{2})\,\delta_a. 
\end{eqnarray}
With the $\delta_c$ needed to detect the quantum states (above), the backaction shifts are large.  The range of excited axial states is given by the size of the average axial quantum number for a Boltzmann distribution,
\begin{equation}
 \bar{n}_z=k_\mathrm{B}T/(\hbar\omega_z) \approx 10   
\end{equation}
for  $\omega_z/(2\pi)=200$ MHz and a detector temperature of $T=0.1$ K. The backaction cyclotron linewidth, greater than $\bar{n}_z \delta_c$, limited the accuracy of past electron measurements. 

Since the magnetic bottle shift $\delta_c$ cannot be reduced without making it impossible to detect the cyclotron and spin state, it is advantageous to find a way around the backaction linewidth.  The cyclotron width produced by the axial ground state ($n_z=0$) is the axial decoherence width $\bar{n}_z \gamma_z$, where $\gamma_z$ is the damping rate for  the axial oscillation energy due to the detection circuit.  The proposal in References \cite{Fan2020EvadingBackActionPRL,Fan2020EvadingBackActionPRA} is to make  $\bar{n}_z\gamma_z$ much less than the cyclotron frequency shift (Eq.~(\ref{eq:CyclotronFrequencyShift})) between the axial ground state and the first excited state, 
\begin{equation}
\bar{n}_z\gamma_z \ll \delta_c
\label{eq:quantizedcondition}
\end{equation}
In this ``strongly dispersive regime" \cite{YaleDispersiveRegime}, the broad cyclotron resonance of width greater than $\bar{n}_z \delta_c$ is resolved into individual resonance lines, each of which corresponds to a particular axial state and quantum number. Measuring the resonance line that corresponds to the axial ground state, with $n_z = 0$, will make it possible to determine the cyclotron frequency that is shifted only by the zero point motion of the axial oscillation (see also Sec.~\ref{Sec:PenningTrapMeasurement}).  

To realize this proposal for circumventing the detector backaction requires that the axial damping caused by the detector $\gamma_z$ be reduced during the time that one-quantum cyclotron transitions are driven.  However, since the induced signal across a detection circuit is proportional to the axial damping rate, this rate must be switched back to a higher value during the time of the measurement that the cyclotron quantum state is read out.  The axial damping rate was $\gamma_z/(2\pi)=1$~Hz in the best measurement \cite{HarvardMagneticMoment2008}, so it requires about two orders of magnitude of reduction to achieve Eq.~(\ref{eq:quantizedcondition}). The rest of this paper is a proposal for switching the axial damping constant, a first demonstration of such switching, and a concluding estimate of the improvement in electron and positron magnetic moment measurement that should now be possible.

\section{Detection Circuitry}
\label{sec:developments}

\subsection{Impedance and Damping}

\begin{figure}
    \centering
    \includegraphics[width=\the\columnwidth]{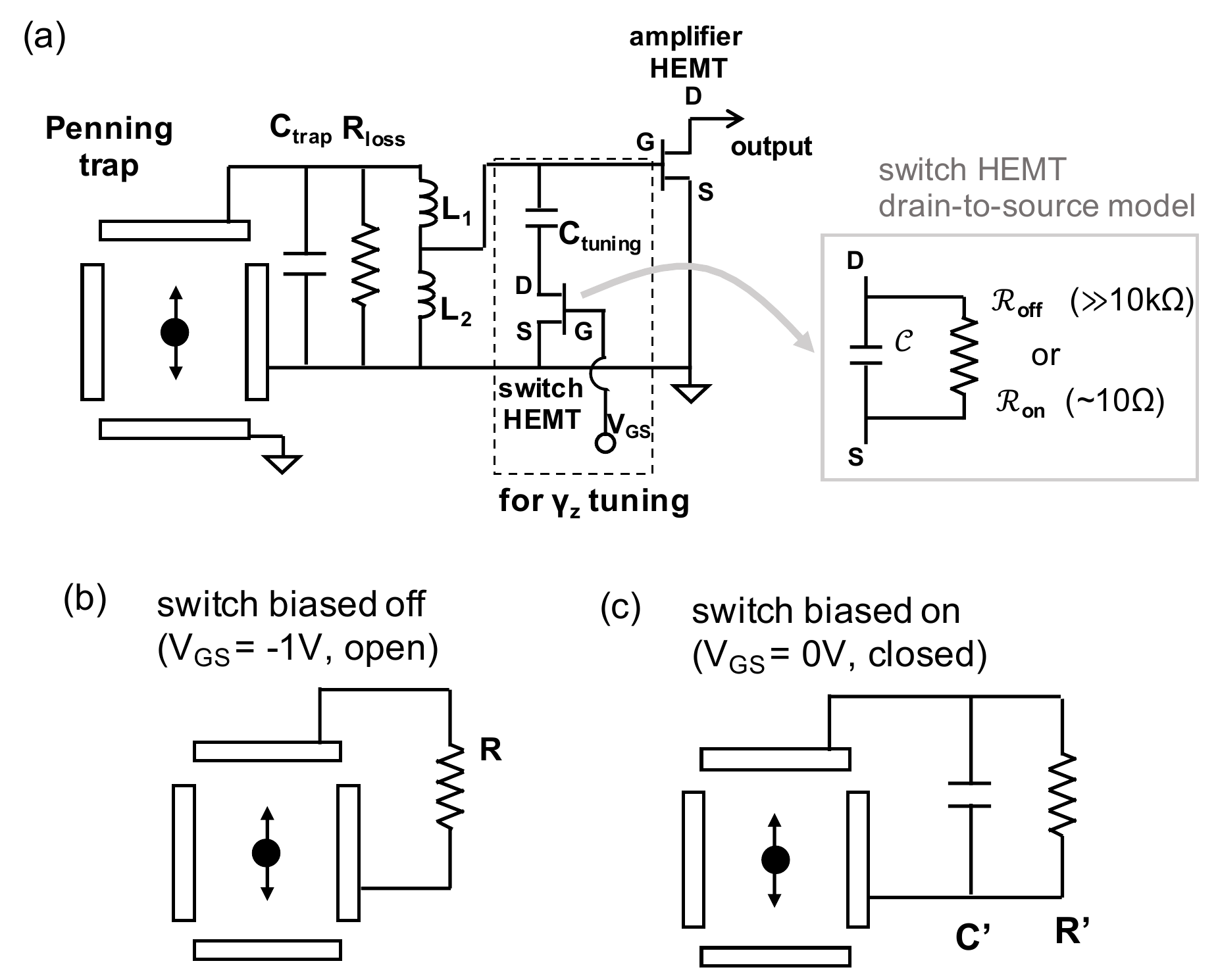}
    \caption{(a) The electron trap and RF electronic circuit used to detect its axial motion. The switch HEMT model is used in Sec.~\ref{sec:developments} D. The electron sees the effective circuits when the HEMT in  dashed box is biased off (b) or on (c). } \label{fig:TrapAndAmp}
\end{figure}

The axial motion of a trapped electron along the direction of a strong magnetic field within surrounding trap electrodes is represented in Fig.~\ref{fig:TrapAndAmp}(a).  This motion induces a 200 MHz electrical current in the frequency dependent impedance $Z(\omega)$ of an attached electrical circuit that can be switched between two circuit states. The electronic switch between the two circuit states is a high electron mobility transistor (HEMT) in series with capacitor $C_\mathrm{tuning}$.  This switch is shown within the dashed rectangle in the figure, and its operation will be described presently (Sec.~\ref{sec:MinimalCoupling}).   

The first state of the circuit  (Fig.~\ref{fig:TrapAndAmp}(b)) is a purely resistive circuit impedance $R$ that pertains at the circuit's resonant frequency, which is made equal to the electron oscillation frequency.  This resistance is due to the unavoidable loss in the circuit and the effective input resistance of the HEMT, rather than from a physical resistor that soldered into the circuit.  The value of $R$ is largest when the losses are the smallest.  The largest possible $R$ is desired because this maximally damps this motion and produces the largest possible oscillating voltage that can be Fourier transformed to determine the axial oscillation frequency (Sec.~\ref{sec:MaximalCoupling}).   

The second state of the circuit is effectively a capacitance and resistance in parallel (Fig~\ref{fig:TrapAndAmp}(c)) that is intended to couple as weakly as possible to the electron motion.  The high frequency current that the electron motion induces in the circuit flows primarily through the capacitor.  Very little signal voltage is produced so that this circuit state is not useful for detection.  Because very little power is dissipated, there is also very little damping of the electron motion. This circuit state is extremely useful for spectroscopy just because the electron motion is so weakly coupled to the circuit that the electron motion is perturbed very little.

With no detection circuit attached, the harmonic oscillation of a particle displaced from equilibrium by $z$ is described by the familiar equation of an undamped harmonic oscillator with  
\begin{equation}
    \frac{d^2z}{dt^2}+\omega_z^2z=0.
    \label{eq:HO}
\end{equation}
The axial oscillation frequency, $\omega_z$, is determined by the DC bias voltages applied to the trap electrodes and by the geometry of the trap electrodes \cite{Review,CylindricalPenningTrap,CylindricalPenningTrapDemonstrated}.  It was at $\omega_z/(2\pi)=200$ MHz in the most recent experiment \cite{HarvardMagneticMoment2011}.
Using complex circuit notation, the real displacement $z$ is given by
\begin{equation}
z=\mathrm{Re}(z[\omega] e^{i\omega t}),     
\end{equation}
where $z[\omega]$ is complex.  
Eq.~(\ref{eq:HO}) becomes $(-\omega^2+\omega_z^2) \,z[\omega] = 0$. The familiar solution is a free oscillation at frequency $\omega_z$ whose fixed amplitude and phase is described by the complex $z[\omega_z]$.

The detection circuit is attached to the trap electrodes as shown in Fig.~\ref{fig:TrapAndAmp}(a). The top electrode is the detection endcap of the trap.  The ring electrode to which the circuit is attached, and other electrodes, are grounded for frequencies at or near the axial frequency, $\omega_z$. The DC lines needed to bias the trap electrodes, and those for biasing the two high electron mobility transistors (FHX13LG from Fujitsu) are omitted because they are designed to not affect the RF behavior of the circuit.  These bias lines for the HEMTs, and all leads entering the cryostat, are carefully filtered to keep external noise from heating the trapped electron.

The oscillation of a particle with charge $e$ induces an oscillatory displacement current that  flows through the electrodes and circuit,
\begin{equation}
   I=\frac{e\kappa}{2z_0}\frac{dz}{dt}, \end{equation}
   where $z_0$ is a trap dimension and $\kappa$ is the fraction of induced charge determined by the trap geometry \cite{CylindricalPenningTrap}.
In complex circuit notation, with $I = \textrm{Re}[I[\omega]e^{i\omega t}]$, 
\begin{equation}
   I[\omega]=\frac{i\omega e\kappa}{2z_0} z[\omega].   \end{equation}
The imaginary number indicates that the current is 90 degrees out of phase with the electron oscillation that induces it.

An oscillatory voltage $V[\omega]$ is induced between the detection endcap and the ring when the induced current flows through the circuit.  This creates an electrical force on the a particle of charge $e$, 
$-e\kappa V[\omega]/(2z_0)$ that opposes the motion. The circuit presents a complex impedance, $Z(\omega)$, across the two trap electrodes to which it is connected,  whereupon   $V[\omega]=I[\omega]Z(\omega)$.  The equation of motion for the particle with mass $m$ and the circuit together is 
\begin{equation}
   \left[ -\omega^2+\omega_z^2
    +i \omega \frac{ Z(\omega)}{m}\left(\frac{e\kappa}{2z_0}\right)^2 
   \right] z[\omega]=\frac{F[\omega]}{m},
\end{equation}
or
\begin{equation}
   \left[ -\omega^2+\omega_z^2
    +i \omega \gamma_z \frac{Z(\omega)}{R} 
   \right] z[\omega]=\frac{F[\omega]}{m},
   \label{eq:ParticleAndCircuit}
\end{equation}
where a driving force $F=\mathrm{Re}[F[\omega] e^{i \omega t}]$ is included, and the constant, 
\begin{equation}
     \gamma_z=
\frac{1}{m}\left(\frac{e\kappa}{2z_0}\right)^2 R
\label{eq.DampingRateZ}   
\end{equation}
is the damping rate of electron motion coupled to a resistive impedance $R$.  We will take $R$ to be the maximal resistive impedance of detection circuit. In Eq.~(\ref{eq:ParticleAndCircuit}), the circuit impedance $Z(\omega)$ is scaled by this maximal resistance. The damping rate for general impedance is given by $\gamma_z\times(\mathrm{Re}[Z(\omega_z)]/R)$, and the suppression parameter
\begin{equation}
\eta=\frac{R}{\mathrm{Re}[Z(\omega_z)]}
\label{eq:eta}
\end{equation}
defines how much the damping is suppressed for general $Z(\omega)$ from its maximum value $R$.


\subsection{Maximal Coupling}
\label{sec:MaximalCoupling}
The maximum coupling of the electron axial motion and the detection circuit takes place when the HEMT switch within the dashed box in Fig.~\ref{fig:TrapAndAmp}(a) is biased ``off," with a gate-to-source voltage of $V_{GS}=-1$ V typically.  The effective impedance of the HEMT and $C_\mathrm{tuning}$ together is then large enough that it can be neglected in determining the impedance that the detection circuit presents to the electron axial motion.

The detection circuit is constructed so that the reactance of the inductor $L = L_1 + L_2$ cancels the reactance of the capacitance of the circuit at the oscillation frequency of the electron oscillation, $\omega_z$.  Near to this frequency, the circuit acts as a pure resistance, $Z(\omega_z)=R$.  When the induced current flows through the resistance, the $I^2R$ loss removes energy from the axial motion of the electron.  Since $Z(\omega_z) = R$, the damping rate from Eqs.~(\ref{eq:ParticleAndCircuit}-\ref{eq.DampingRateZ}) is $\gamma_z$.  Experiments have shown that $R>60$ k$\Omega$ suffices so that the one-particle signal can be Fourier transformed to ascertain that oscillation frequency\cite{HarvardMagneticMoment2008}. Shifts in this measured frequency signal one-quantum transitions of the spin and cyclotron motions\cite{QuantumCyclotron,Fan2020EvadingBackActionPRL}.  

If there is no driving force (i.e.\ $F=0$), an initial displacement $\tilde{z}_0$ of the particle from equilibrium damps exponentially as   
\begin{equation}
z = \mathrm{Re}\left[\tilde{z}_0\,e^{i \omega_z t}\right] e^{-\tfrac{1}{2}\gamma_z t}. \end{equation}
The amplitude and energy of this free oscillation damp with time constants $2/\gamma_z$ and $1/\gamma_z$.  
A resonant driving force $F_0\cos\left(\omega_zt\right)$ applied for a time 
$t \gg (\gamma_z)^{-1}$
sets up a steady state 
\begin{equation}
z = \frac{F_0}{m\gamma_z\omega_z} \sin{(\omega_z t)}.
\end{equation}
The driving force equals the damping force for this steady state.  The steady-state signal, because it does not damp out, can be averaged to determine the electron's axial oscillation frequency as accurately as needed.  In principle, increasing the drive force $F$ makes the amplitude of the driven particle to be arbitrarily large.  In practice, the oscillation amplitude and hence the particle velocity must be kept small enough so that the particle oscillation remains in the harmonic potential region near the center of the trap.

The maximum power dissipated by the current flowing through the resonant detection circuit is
\begin{equation}
P=m \gamma_z \dot{z}^2
\label{eq:SignalPower}
\end{equation}
For constant oscillation amplitude, $P$ is proportional to $\gamma_z$. The cryogenic detection circuit is designed to make $R$ as large as possible, thus maximizing both the damping rate $\gamma_z$ and the signal power.  The circuit with its HEMT amplifier is designed to maximize the signal power sent out of the dewar to be Fourier-transformed to determine the electron axial oscillation frequency $\omega_z$.  

The detection circuit resonant at the electron oscillation frequency acts as a simple resistance $R$ through which the induced current flows. This current would primarily flow through the capacitance between the trap electrodes, $C_\mathrm{trap}=8.2$ pF except that a parallel inductor $L = L_1 + L_2 = 70$ nH cancels the reactance of the capacitor to prevent this. (The cancelled capacitance also includes small contributions from the distributed capacitance in the inductor, and from the input capacitance of the amplifier HEMT that is outside the dashed box in the circuit figure.)  The inductor is tapped such that $L_1=55$ nH and $L_2=15$ nH to match the impedance of the tuned circuit and HEMT amplifier.  
The tapped inductor is a $78~\Omega$ coaxial transmission line resonator, tapped at 15 cm out of its total length of 20 cm. The mutual inductance between $L_1$ and $L_2$ is negligible in such a resonator. 

The resistance $R$ is due to the RF losses that cannot be avoided in the circuit, $R_\mathrm{loss}$, along with a contribution from the input impedance of the amplifier HEMT and the switching circuit. No explicit resistive element is added to the circuit. Since $R = Q \omega_z L$, the effective value of this resistance is determined by the measured resonance frequency, $\omega_z$, the measured inductance $L = L_1+L_2,$ and the quality factor of the $LCR$ circuit that is determined by the observed resonance width.  For these demonstration experiments, we typically observed $Q=800$ and $R = 83$ k$\Omega$.  
 
\subsection{Minimal Coupling}
\label{sec:MinimalCoupling}

While one-quantum cyclotron and spin transitions are being driven to make a magnetic moment measurement, the high $R$ and high $\gamma_z$ are not desirable. These would cause the backaction of the strongly coupled detector upon the electron. As a result, the electron's axial amplitude fluctuates within a magnetic field gradient, causing fluctuations of the electron cyclotron and spin frequencies. 

What is desirable during the time that these quantum transitions are being  driven is the smallest possible circuit resistance and axial damping rate, $\gamma_z$. 
Heater or varactor based tuners have been used previously in different environments\cite{BaseDetector2016Nagahama}.
However, in a 0.1 K apparatus at 5 Tesla, switching the resistance of the circuit from a high value to a low one is not a trivial task. 
The solution explored here is the use of an electrical HEMT switch 
(within the dashed box in Fig.~\ref{fig:TrapAndAmp}(a)) that can be switched rapidly and reliably with the low heat dissipation needed to operate at 0.1 K. 

A HEMT gate-to-source voltage of $V_{GS}=0$ V makes the HEMT act like a small resistance.  It is in series with a capacitor called $C_\mathrm{tuning}$. Its major effect is tuning the detection circuit's resonant frequency to a much lower value. For example, for a typical value $C_\mathrm{tuning}=22$~pF, $\omega_0^\prime/(2\pi)=172$ MHz, with a parallel resistance $R^\prime=3.4$ k$\Omega$ and quality factor $Q^\prime=45$. The current induced by the electron oscillation at $\omega_z$ thus sees an effective detection circuit that is a resistor $R^\prime$ in parallel to a capacitor $C^\prime$ (Fig.~\ref{fig:TrapAndAmp}(c)).  The effective damping resistance $\mathrm{Re}[Z(\omega_z)]$ is reduced by approximately
$300$ and the damping rate $\gamma_z$ is reduced by a factor of $\eta=R/\textrm{Re}[Z(\omega_z)] \approx 300$. In Sec.~\ref{sec:optimization} we experimentally determine the parameters of this effective circuit.

\subsection{Optimization of the Tuning Capacitor}
\label{sec:optimization}

The value of the capacitor $C_\mathrm{tuning}$ must be experimentally optimized, since at 200 MHz the stray capacitance and inductance modify the nominal component values.  When the HEMT switch is biased off, $C_\mathrm{tuning}$ must be small enough so that its reactance keeps the biased-off HEMT drain-to source resistance from unacceptably lowering the $R$, $\gamma_z$ and the detection sensitivity.  At the same time, $C_\mathrm{tuning}$ must be large enough that its reactance and the low drain-to-source resistance of the biased-on HEMT will reduce $\gamma_z$ as much as possible.  

\begin{figure}
    \centering
    \includegraphics[width=\the\columnwidth]{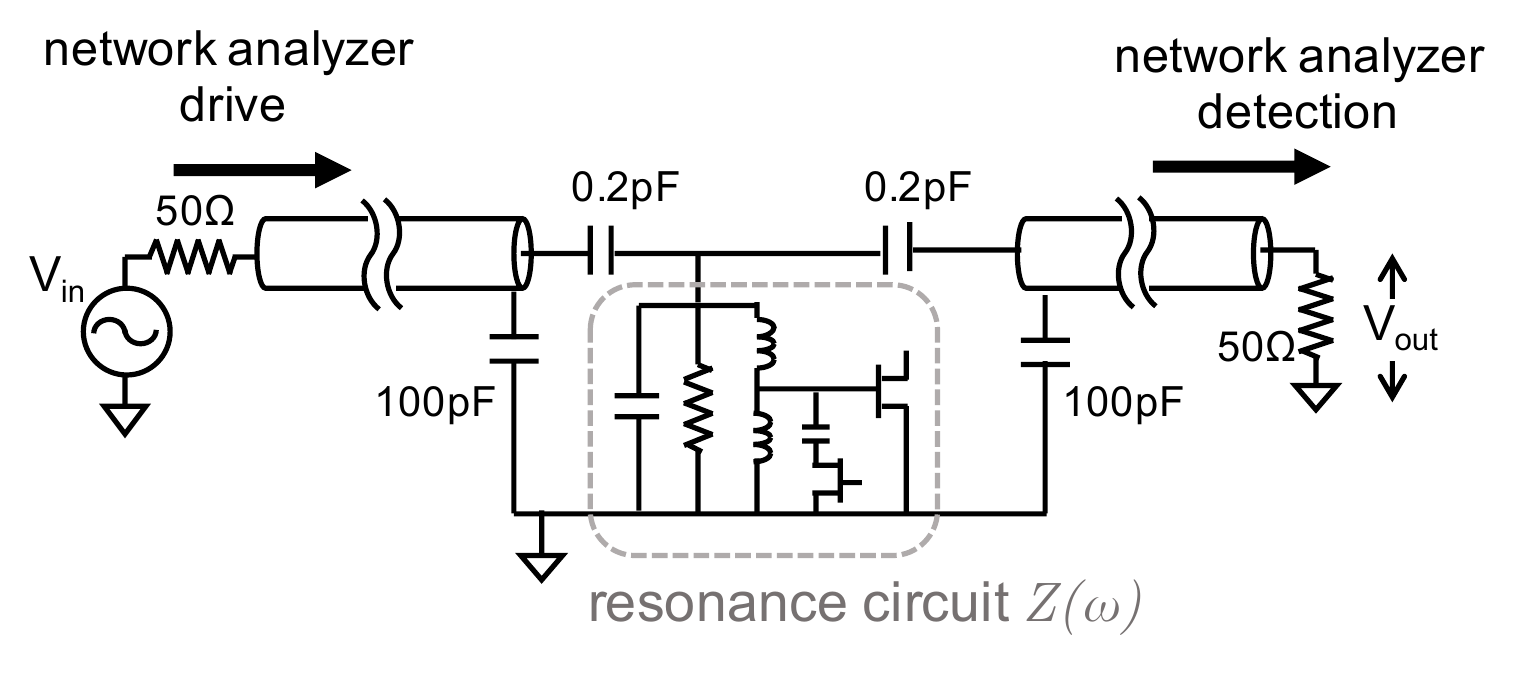}
    \caption{Circuit to measure the impedance $Z(\omega)$ of the resonance detection circuit} \label{fig:ImpedanceMeasurement}
\end{figure}

With a low loss capacitor substituted for trap electrodes for the optimization measurements, the effective resistance on resonance was about 83 k$\Omega$. The circuit was located in a dewar that could be cooled as low as 3.1 K by a pulse tube refrigerator in a 0 to 6 Tesla solenoid magnet. A network analyzer injects a small drive into the dewar and circuit through the weak coupling of a capacitive divider made with 0.2 pF and 100 pF capacitors (Fig. \ref{fig:ImpedanceMeasurement}). The detector of the network analyzer is similarly weakly coupled to the circuit. The weak couplings increase the impedance of the 50 $\Omega$ input and output lines to $Z_0^\prime\approx\left(100~\mathrm{pF}/0.2~\mathrm{pF}\right)^2\times 50~\Omega=12.5~\mathrm{M}\Omega$. The loss and phase shift of the cables are calibrated by shorting them by a straight connector. As in the 2 port shunt-through measurement \cite{ShuntThru}, when $|Z(\omega)|\ll Z_0^\prime$, the transmission $V_\mathrm{out}/V_\mathrm{in}$ is a function of the impedance of the resonance circuit $Z(\omega)$ as
\begin{equation}
    \frac{V_\mathrm{out}}{V_\mathrm{in}}=\frac{Z(\omega)}{Z_0^\prime+2Z(\omega)}.
    \label{eq:ShuntThru}
\end{equation}
Thus, the impedance $Z(\omega)$ can be calculated from the transmission amplitude and phase.

\begin{figure}
\centering
\subfloat{\includegraphics[width=\the\columnwidth]{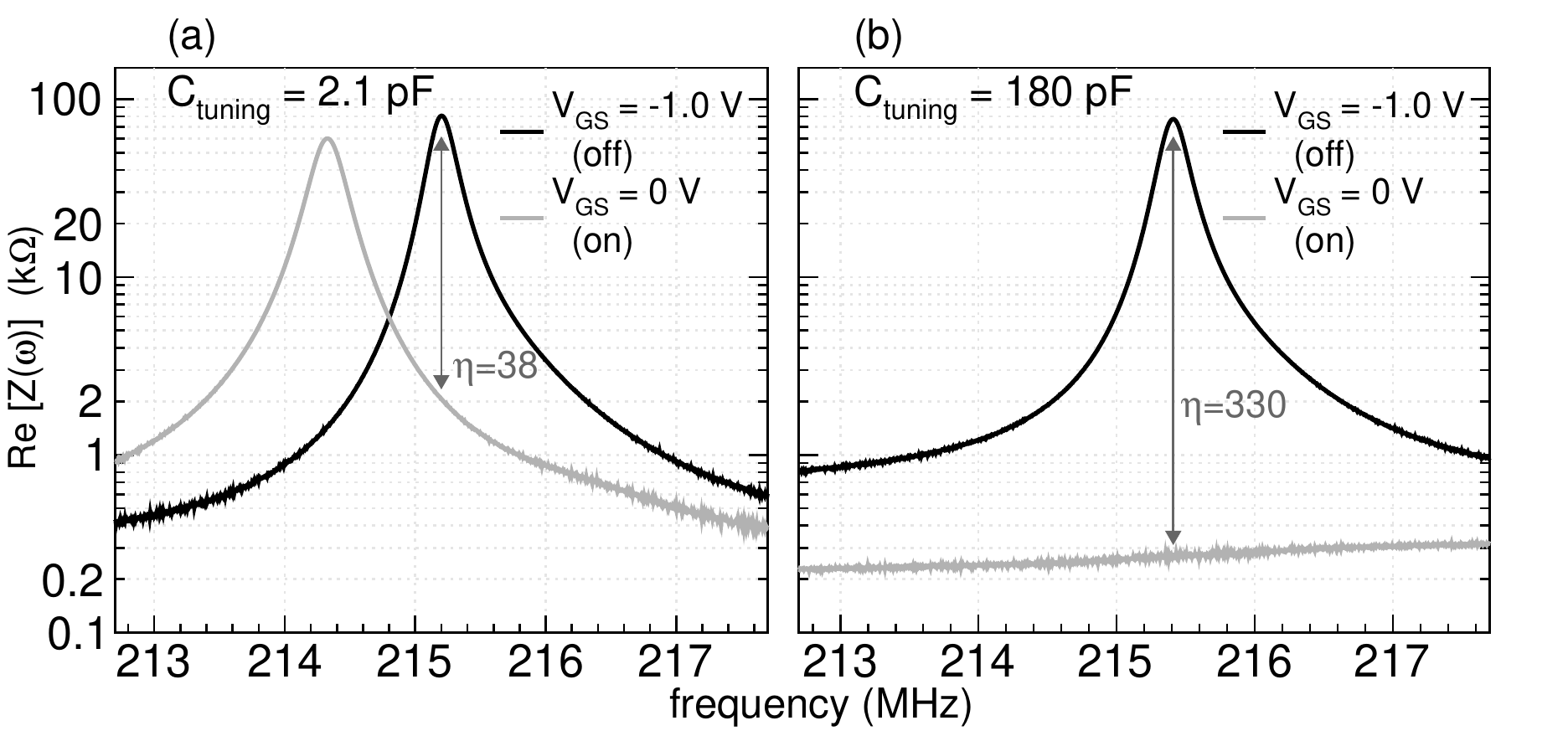}}
\hfill
\subfloat{\includegraphics[width=\the\columnwidth]{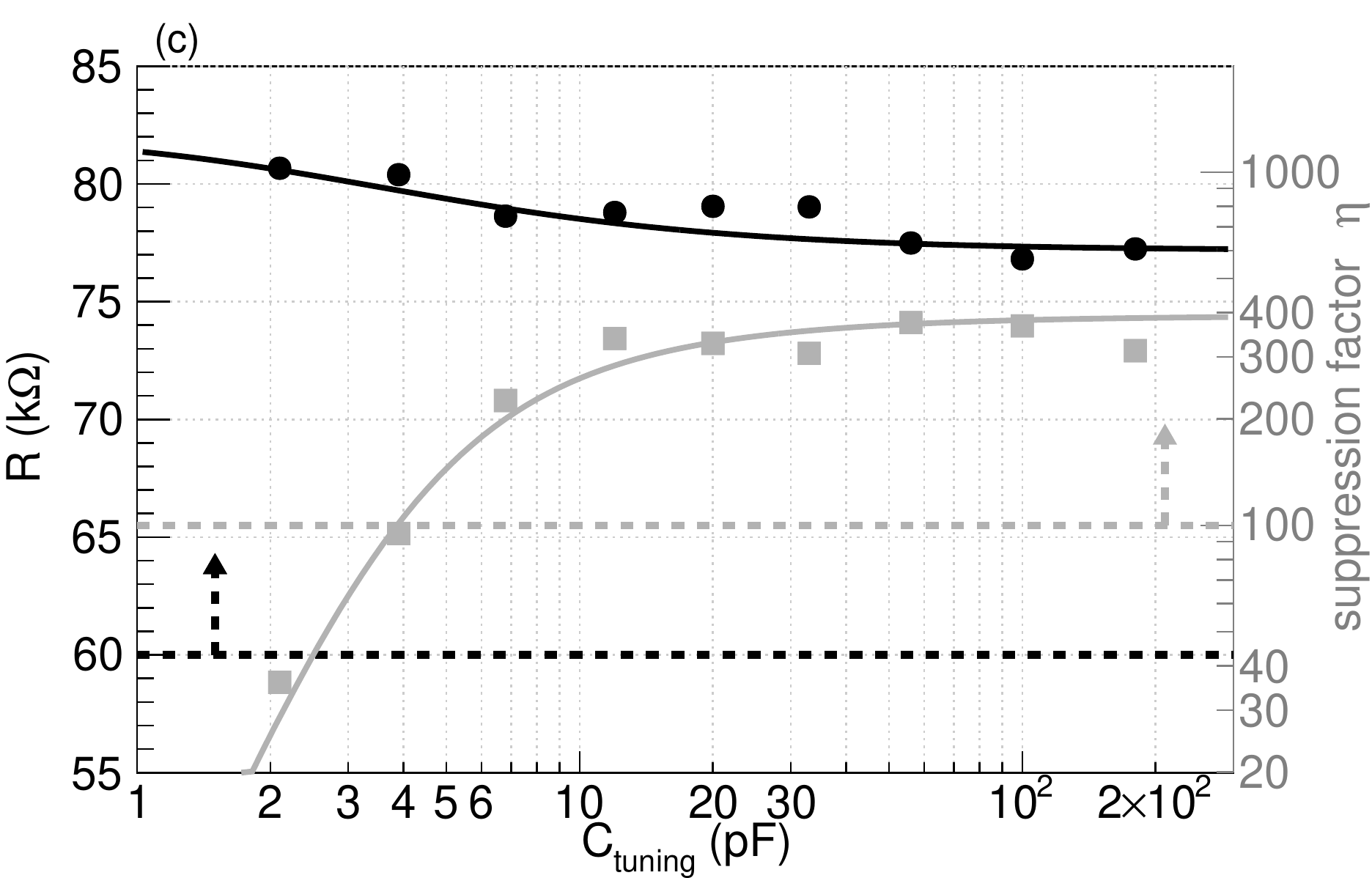}}
\caption{Measured $\mathrm{Re}[Z(\omega_z)]$ for a transmission line resonator
with the HEMT switch on and off, for $C_\textrm{tuning}$ values of 2.1 pF (a) and 180 pF (b). (c) For various $C_\textrm{tuning}$, the measured parallel $R$ for the HEMT switched off are the black dots, with a higher values desired. The measured reduction of the damping resistance as the HEMT is switched on is shown in gray, with the largest value desired. The dashed lines shows the requirements for $R$ and $\eta$ respectively.} 
\label{fig:ctuningoptimization}
\end{figure}

Figure \ref{fig:ctuningoptimization} shows the optimization for $T=3.1$ K and 5 T. $Z(\omega)$ is measured from Eq.~(\ref{eq:ShuntThru}). The reduction of the damping resistance as the HEMT is switched on, at the frequency $\omega_z = (2\pi)\times215.3$ MHz, is given by Eq.~(\ref{eq:eta}) with $R=\mathrm{Re}[Z(\omega_z)]_\mathrm{off}$,
\begin{equation}
\eta=\frac{\mathrm{Re}[Z(\omega_z)]_\mathrm{off}}{\mathrm{Re}[Z(\omega_z)]_\mathrm{on}}.
\end{equation}
For a small $C_\mathrm{tuning}=2.1$ pF in (a), the resonance shifts slightly downward as the HEMT is switched on, and the damping of an electron oscillation at 215.3 MHz is reduced by $\eta=38$.  For a large $C_\mathrm{tuning}=180$ pF in (b), the resonance shifts downward much more when the HEMT is switched on, and the damping of an electron oscillation at 215.3 MHz is reduced by $\eta=330$. In this case, most of the current induced by an electron oscillation at this frequency goes though the capacitance in Fig. \ref{fig:TrapAndAmp}(c) rather than through a damping resistance. 

The off-state resonance shapes for these value of $C_\mathrm{tuning}$ and others shown in (c) are fit to a Lorentzian to get the quality factor $Q$  needed to deduce the $R=Q\omega_z L$ (black dots in (c)). The gray dots in (c) are the reduction factor $\eta$ determined from traces like those as in (a) and (b). 

The impedance between the detection electrode and ground can be analytically calculated using the electronics model in Fig.~\ref{fig:TrapAndAmp} (a). $C_\mathrm{trap}$, $L_1$, $L_2$ are independently determined from the resonance shape and the dimension, and $C_\mathrm{tuning}$ is a controlled variable here. The drain-to-source impedance of the HEMT is modeled by a capacitor $\mathcal{C}$ and a resistor $\mathcal{R}_\mathrm{off}$ or $\mathcal{R}_\mathrm{on}$ in parallel (depending on $V_{GS}$, thus the switch state). We can calculate theoretical $R$ and $\eta$ from this model for these given parameters.
The measured $R$ and $\eta$ in Fig.~\ref{fig:ctuningoptimization} (c) are fitted by this model. As discussed above, the only free parameters are $\mathcal{C}$, $\mathcal{R}_\mathrm{off}$, and $\mathcal{R}_\mathrm{on}$. The best fit values are $\mathcal{R}_\mathrm{off}=65~\mathrm{k}\Omega$, $\mathcal{R}_\mathrm{on}=9.6~\Omega$, and $\mathcal{C}=1.8$ pF.

As discussed earlier, when the HEMT switch is biased off, a large $R$ is needed to effectively detect quantum transitions made by a trapped electron, and to effectively damp the electron's axial motion.  Past experiments have demonstrated that $60~\mathrm{k}\Omega$ suffices. When the switch is biased on, a reduction of the axial damping rate by a factor of $\eta=100$ (or greater) is desired from Eq.~(\ref{eq:quantizedcondition}). The $R$ value stays quite high over most of the range of capacitor values used.  The factor $\eta$ increases to a saturation value as $C_\mathrm{tuning}$ increases, leading us to choose $C_\textrm{tuning}=22$ pF for experiments going forward with this detection circuit.

\section{Demonstration and Implications}
\label{Sec:PenningTrapMeasurement}

A switchable detection circuit attached to a Penning trap with a 5.3 Tesla field provides the first demonstration of switching the damping resistance the circuit presents to the electrons. The large signal from the center-of-mass motion of order of a thousand electrons is used to demonstrate that the circuit can make the desired damping resistance changes.  The Penning trap and the detector system are cooled to an ambient temperature of 0.1 K by a dilution refrigerator \cite{atomsNewMeasurement2019}. The effective circuit resistance $R$ is thereby kept as low in temperature as possible to minimize the thermal Johnson noise in the resistor that heats the electron at the same time as it damps its 200 MHz motion.  Cold damping is important for both a high frequency electron oscillator  and low frequency oscillator with a large mass \cite{ColdDamping1,ColdDamping2}. To this end, the amplifier HEMT used here is operated with a power dissipation of only about 120 $\mu$W that still heats the circuit to a temperature of 5--10 K. 

\begin{figure}
    \centering
    \includegraphics[width=\the\columnwidth]{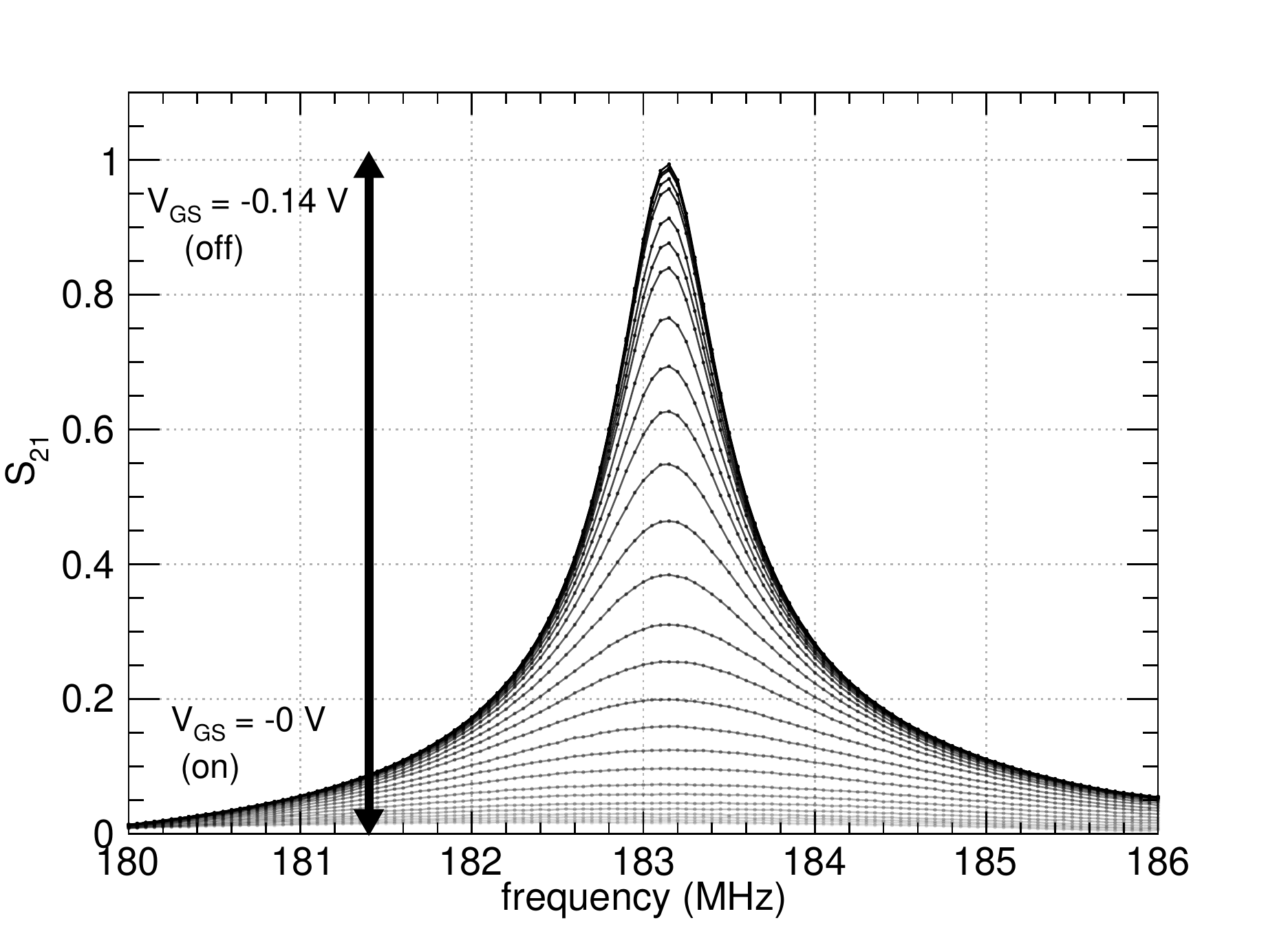}
    \caption{Demonstration of the tuning of the quality factor of a detection circuit connected to a Penning trap. The top curve is for the HEMT biased off (with a gate-to-source bias at $V_{GS}=-0.14$ V). Each line corresponds to different $S_{21}$ with $V_{GS}$ reduced by 0.005 V in each step.}  
    \label{fig:Qtuning}
\end{figure}

The resistance $R$ used for this demonstration was only $36~\mathrm{k}\Omega$ for unrelated reasons related to loading positrons into the trap. The behavior of the switchable detection circuit was characterized by applying an RF drive to another electrode and measuring the transmitted amplitude from the detector HEMT, $S_{21}$, the well known $S$ parameter that characterizes the transmission of a signal amplitude through the circuit. Figure \ref{fig:Qtuning} shows $S_{21}$ as a function of gate-source bias voltage, $V_{GS}$, on the switch HEMT. No particle is in the trap in this measurement. Since only the relative change of $S_{21}$ is meaningful, the value is scaled so that the peak value is 1. The effective resistance $R$ can be tuned from $36~\mathrm{k}\Omega$ to around 100 $\Omega$. 

\begin{figure}
    \centering
    \includegraphics[width=\the\columnwidth]{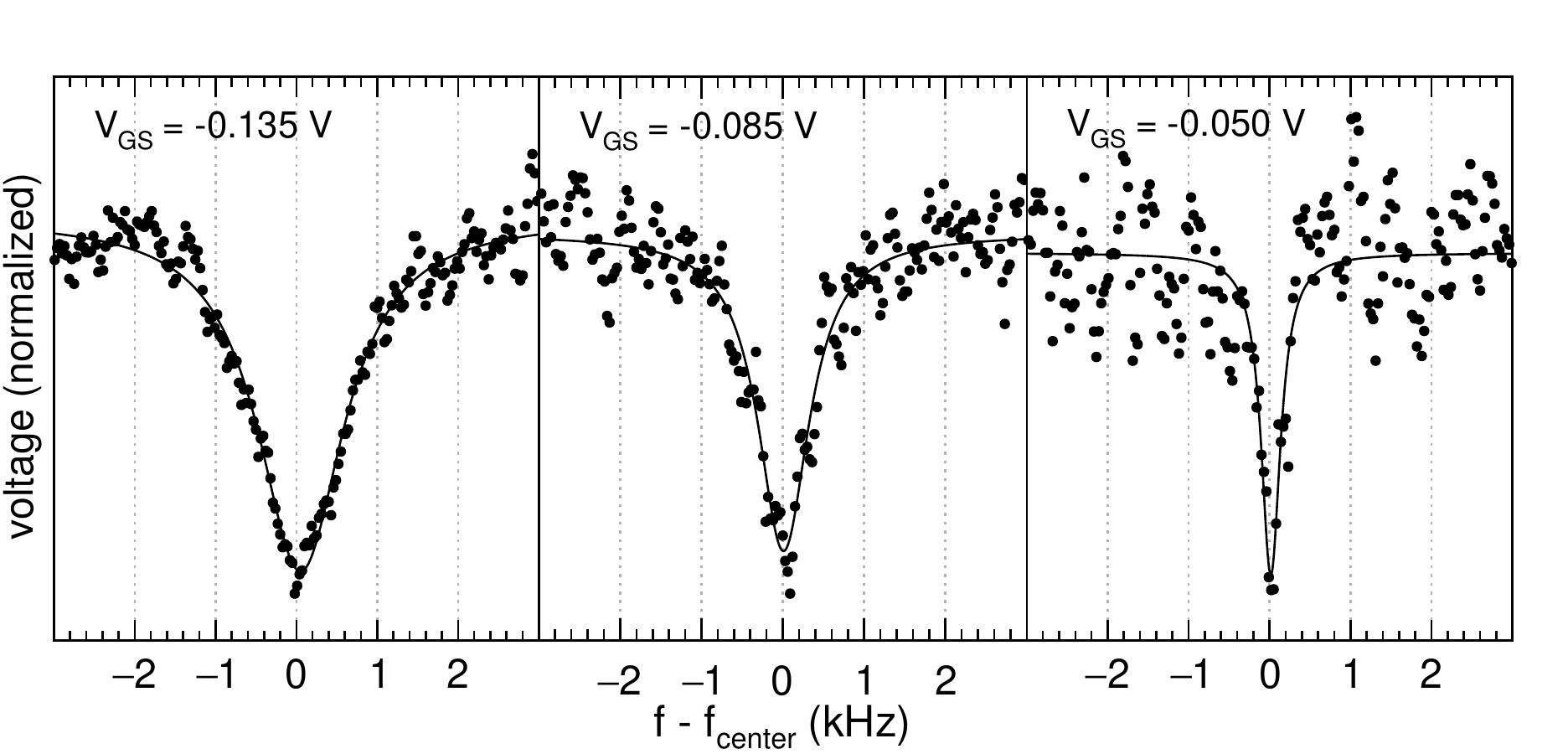}
    \caption{Change of the dip width of a cloud with $N\approx 1500$ particles. The gate voltage of the switch HEMT $V_{GS}$ is adjusted to set the $R$ to be about $36~\mathrm{k}\Omega$, $19~\mathrm{k}\Omega$, and $7.6~\mathrm{k}\Omega$ from left to right respectively. The change of the dip width reveals the change oft the electron damping rate. The noise becomes relatively larger for smaller $R$ since the detection sensitivity is also proportional to $R$.}
\label{fig:dipwidth}
\end{figure}

This fine control of the resistance $R$ makes it possible to observe the change of the axial damping rate $\gamma_z$ with the trapped particles (Eq.~(\ref{eq.DampingRateZ})). When a cloud of $N$ electrons is trapped, a dip appears at the axial frequency $\omega_z/(2\pi)$ on the noise resonance \cite{Review}. The full-width-at-half-maximum (FWHM) of the dip is $N\gamma_z$, the damping rate that we seek to be able to change.  Figure~\ref{fig:dipwidth} illustrates the change of the dip width for the center-of-mass motion of about $N\approx 1500$ electrons. The particles are trapped in a well-tuned harmonic trap and their  magnetron motion \cite{Review} is cooled to center them in the trap. Since the amplitude of the detected signal is also proportional to $R$, the signals are smaller for lower $R$. The baseline voltages in the graphs are shifted so that the baseline voltages are the same among three. The measured damping width can then be compared to the  $R$ deduced from the $S_{21}$ measurements in Fig.~\ref{fig:Qtuning} and $R=Q\omega_zL$. 
$R$ is about $36~\mathrm{k}\Omega$, $19~\mathrm{k}\Omega$, and $7.6~\mathrm{k}\Omega$ from left to right, respectively. The damping rate $\gamma_z$ clearly reduces as $R$ is reduced.  


\begin{figure}
    \centering
    \includegraphics[width=\the\columnwidth]{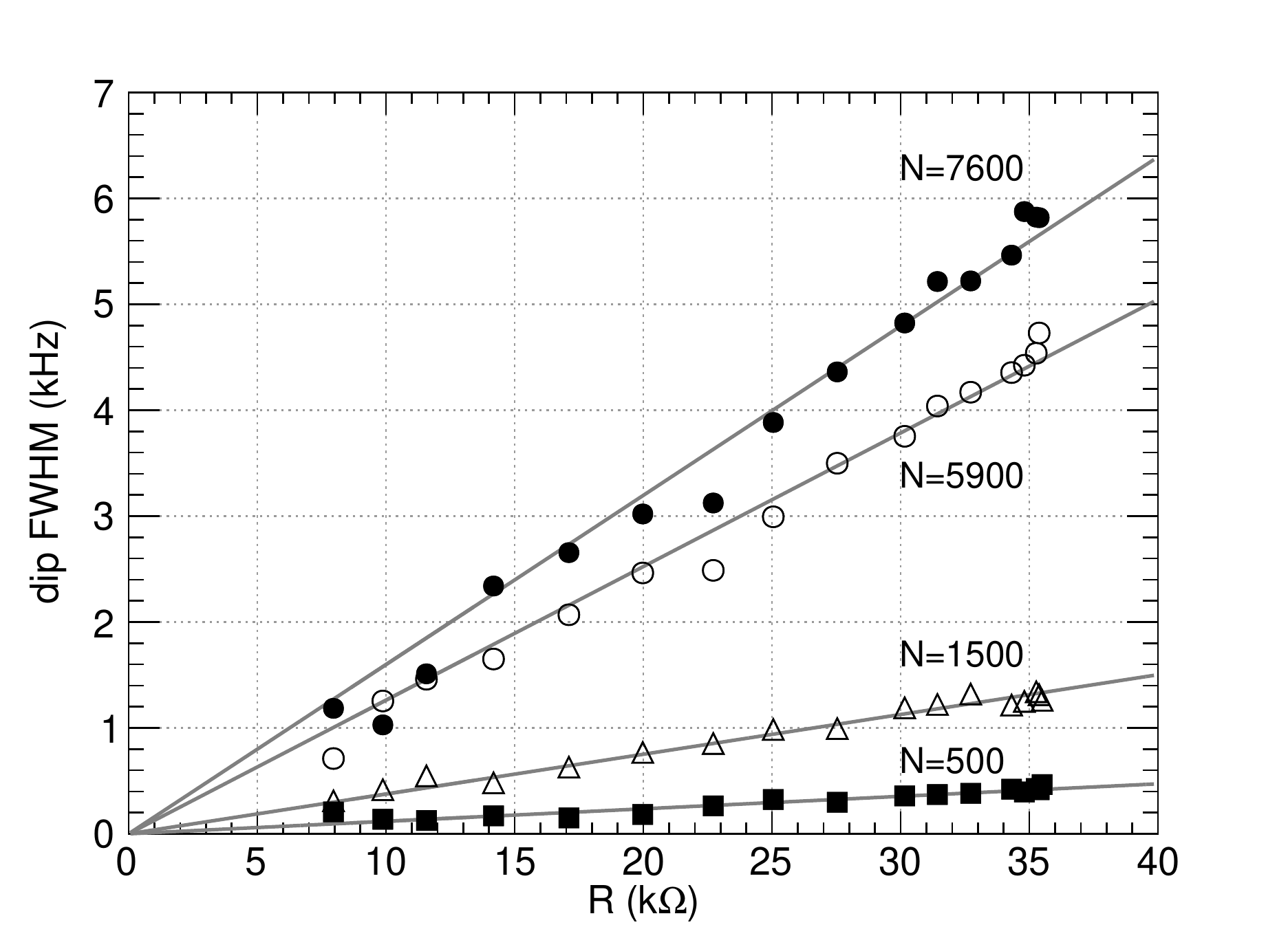}
    \caption{Change of dip width of 4 different sizes of clouds. The $R$ is set by adjusting the gate voltage on the switch HEMT $V_G$. Linear fittings on each cloud are also shown.}
    \label{fig:Qgamma}
\end{figure}
Figure~\ref{fig:Qgamma} summarizes the change of dip widths for four different sizes of clouds. $R$ is set by adjusting the gate voltage $V_{GS}$ on the switch HEMT as in Fig.~\ref{fig:Qtuning}. The data with $R$ lower than $7~\mathrm{k}\Omega$ cannot be detected reliably because the noise resonance is too small to observe the dips. All data are taken with the same trap voltages. Results of linear fitting on each cloud is also shown in the graph. The ability of tuning $\gamma_z$ by changing the $R$ is demonstrated.

The tuning of axial damping rate $\gamma_z$ demonstrates that the HEMT switch is compatible with the Penning trap. The measurements with coaxial transmission line resonator (Sec.~\ref{sec:developments}) shows that the HEMT based switch has low enough loss to detect a single particle and high enough suppression on $\gamma_z$. The suppression of $\gamma_z$ is demonstrated with trapped electrons. With these demonstrations, the newly developed detector is able to reduce $\gamma_z$ enough while maintaining single-particle-detection sensitivity.

\begin{figure}
    \centering
    \includegraphics[width=\the\columnwidth]{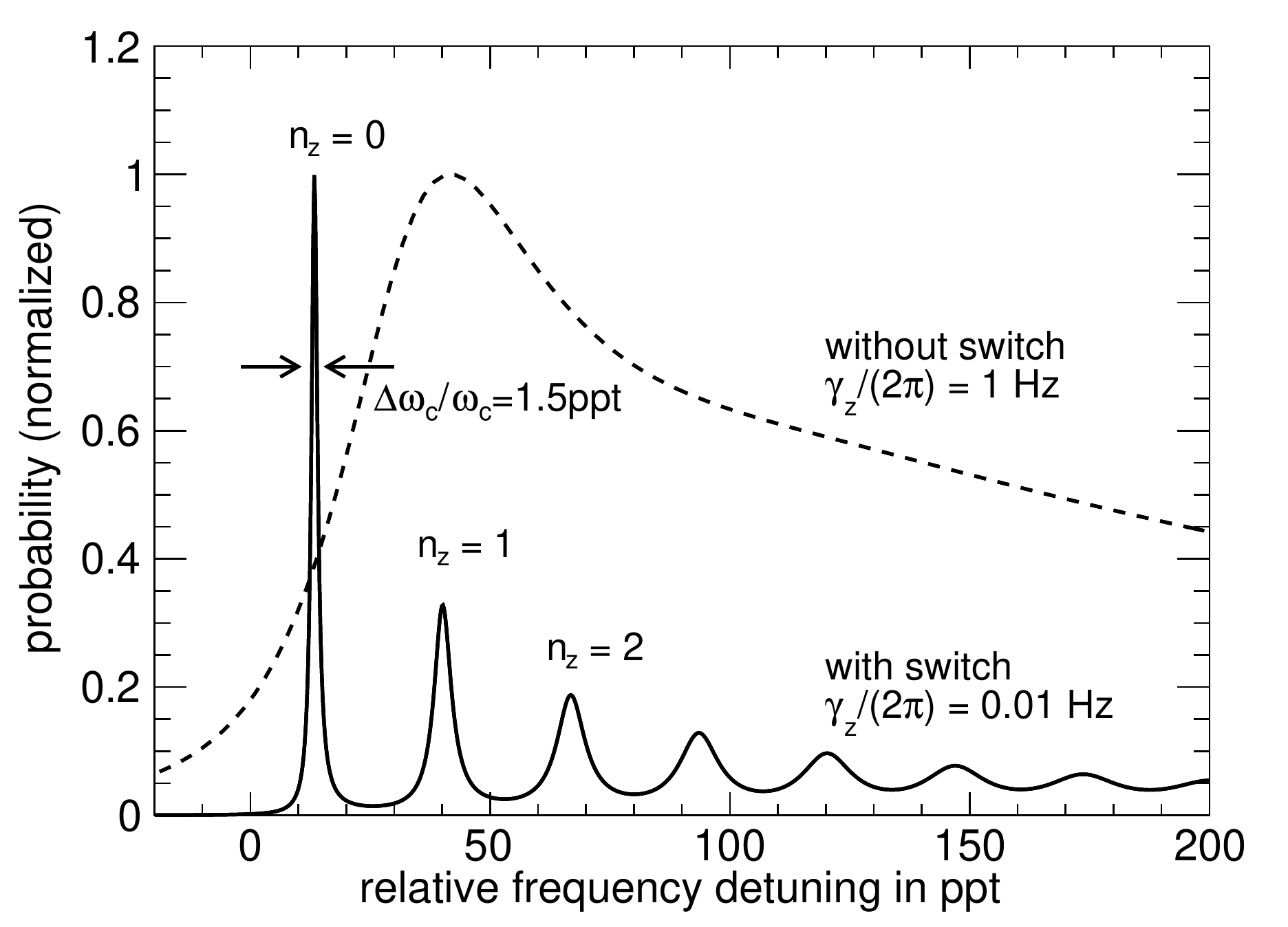}
    \caption{Lineshape of cyclotron transition for traditional measurement without the switch (dashed) and with the demonstrated switch  (solid) for $\bar{n}_z=10$ and $\delta_c/(2\pi)=4$~Hz\cite{HarvardMagneticMoment2008}. See \cite{Fan2020EvadingBackActionPRL} for the details of the calculation.} \label{fig:quantumtransition}
\end{figure}

According to recent calculations \cite{Fan2020EvadingBackActionPRL}, the demonstrated switchable detection circuit should dramatically change the cyclotron resonance lineshape that must be observed to measure the electron and positron magnetic moments. The dashed lineshape is what would be observed if the detection circuit was not switched, as in completed experiments \cite{HarvardMagneticMoment2006,HarvardMagneticMoment2008}.  The dramatically different series of resonances (solid) is what is expected when the detection circuit demonstrated here is switched on.  The broad and asymmetric resonance (dashed) turns into a series of extremely narrow and symmetric peaks, each of which corresponds to an individual quantum state of the axial motion. The linewidth is reduced by about two orders of magnitude. The details of how the detection circuit is used to observe cyclotron resonance is well beyond the scope of this report and is discussed in \cite{Fan2020EvadingBackActionPRL}.  

\section{Conclusions}
\label{sec:Conclusion}

A 200 MHz detection circuit that can be switched between high and low resistive impedance levels has been developed for use at cryogenic temperatures as low as 0.1 K.  The switchable detection and damping circuit is demonstrated by using it to change the damping rate for the axial, center-of-mass motion of trapped electrons. The change in the damping rate for a single electron will be about a factor of 100 for the demonstrated circuit.  According to a recent calculation, being able to switch the damping rate by this factor will make it possible to evade the detector backaction that limited the accuracy of earlier measurements by producing broad and asymmetric cyclotron resonances. The switchable detection circuit thus promises to revolutionize electron and positron magnetic moment measurements made to test the most precise predictions of the standard model of particle physics.

\section*{acknowledgements}
This work was supported by the NSF, with X. Fan being partially supported by the Masason Foundation.

\section*{Data Availability}
The data that support the findings of this study are available from the corresponding author upon reasonable request.

\bibliographystyle{prsty_gg}
\bibliography{ggrefs2020,NewRefs}

\end{document}